\documentclass[journal=jcisd8,manuscript=article]{achemso}

\usepackage[T1]{fontenc}  
\usepackage[numbers]{natbib}  
\usepackage{booktabs}       

\author{Christine H. Chang}
\affiliation[PNNL]{Biological Sciences Division, Pacific Northwest National Laboratory, Seattle, WA 98109}
\email{christine.chang@pnnl.gov}

\author{Bryan J. Killinger}
\affiliation[PNNL]{Biological Sciences Division, Pacific Northwest National Laboratory, Richland, WA 99352}

\author{Ryan S. Renslow}
\affiliation[PNNL]
{Biological Sciences Division, Pacific Northwest National Laboratory, Richland, WA 99352}
\alsoaffiliation[WSU]
{The Gene and Linda Voiland School of Chemical Engineering and Bioengineering, Washington State University, Pullman, WA 99164}

\author{Sean M. Colby}
\affiliation[PNNL]{Biological Sciences Division, Pacific Northwest National Laboratory, Richland, WA 99352}
\email{sean.colby@pnnl.gov}

\title{Identifying metabolites from protein identifiers with P2M}

\abbreviations{EC, UP, SPARQL, ChEBI, SMILES, InChI}

\begin{document}

\begin{tocentry}
\includegraphics[]{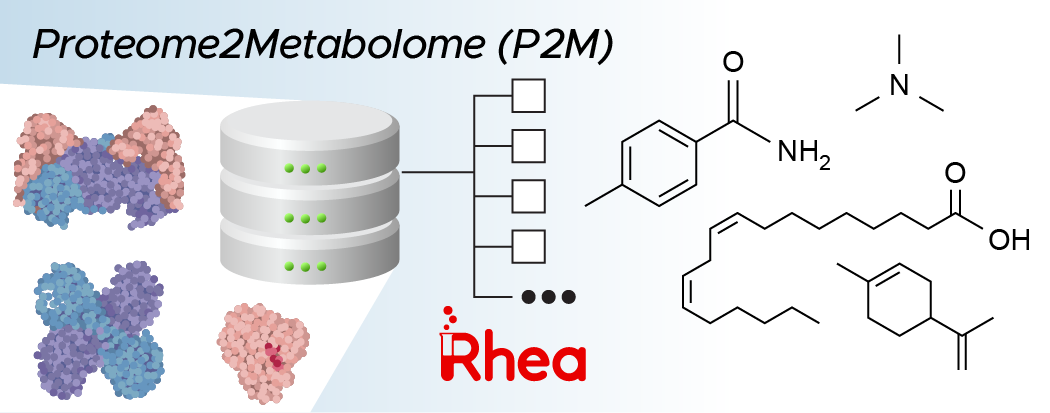}
\end{tocentry}

\begin{abstract}
    The identification of metabolites from complex biological samples often involves matching experimental mass spectrometry data to signatures of compounds derived from massive chemical databases. However, misidentifications may result due to the complexity of potential chemical space that leads to databases containing compounds with nearly identical structures. Prior knowledge of compounds that may be enzymatically consumed or produced by an organism can help reduce misidentifications by restricting initial database searching to compounds that are likely to be present in a biological system. While databases such as UniProt allow for the identification of small molecules that may be consumed or generated by enzymes encoded in an organism's genome, currently no tool exists for identifying SMILES strings of metabolites associated with protein identifiers and expanding R-containing substructures to fully defined, biologically relevant chemical structures. Here we present Proteome2Metabolome (P2M), a tool that performs these tasks using external database querying behind a simple command line interface. Beyond mass spectrometry based applications, P2M can be generally used to identify biologically relevant chemical structures likely to be observed in a biological system.
\end{abstract}
 
\section{Introduction}

Predicting and annotating protein coding sequences are essential to determining the functional capabilities of an organism. With the development of databases linking protein sequences to their associated enzymatic reactions, the ability to classify the functional capacity of organisms is rapidly increasing.\citep{Alcantara2012, Uniprot2015, Bansal2022, Morgat2019} In addition to environmental perturbations, the metabolic profiles of different organisms can vary substantially due to genomic differences, resulting in differential expression of proteins responsible for performing enzymatic reactions. Thus, parallel to sequence-based analysis, the detection of metabolites with high-throughput mass spectrometry has proven to be a valuable tool for investigating the physiological state of an organism.\citep{Dettmer2007} 

While advances in mass spectrometry have continued to increase the accuracy of experimental characterization of metabolites,\citep{Kind2010, Levy2019} mapping mass spectrometry signatures to a chemical identity from complex samples can require searching large chemical databases with potential misidentifications due to spectral similarity between chemicals of similar structure. Distinguishing between spectral signatures of similar compounds is especially critical in untargeted and standards-free identification-based approaches.\citep{Nunez2019} Thus, reducing the initial search space of compounds is valuable for reducing potential incorrect assignments of chemical identity in addition to computational requirements.

Here, we present Proteome2Metabolome (P2M), a tool for identifying a reasonable set of metabolites given an annotated proteome by mapping protein identifiers to enzymatic reactions and their associated substrates and products. Given a list of Universal Protein Resource (UniProt)\citep{Uniprot2015, Morgat2019} or Enzyme Commission (EC)\citep{Committee1992, McDonald2022} protein identifiers, P2M identifies associated Rhea reactions\citep{Alcantara2012, Bansal2022} from which corresponding substrates and products are reported via SMILES string representations.\citep{Weininger1988} In addition to reducing initial mass spectrometry search space, P2M can be used to easily identify if select metabolites are likely to be consumed or produced by an organism. Thus, by initially matching experimental data to metabolites likely to be observed in a biological system, P2M is designed to facilitate the identification of metabolites in complex biological samples.

\section{Methods}

P2M accesses the UniProt and Rhea SPARQL endpoints to: (i) identify enzymatic reactions in the Rhea database associated with protein identifiers and (ii) map substrates and products of enzymatic reactions as SMILES strings. While Chemical Entities of Biological Interest (ChEBI)\citep{Degtyarenko2007, Hastings2016} identifiers associated with UniProt IDs annotated with Rhea reactions can be obtained directly from UniProt, linking an arbitrary list of UniProt identifiers to a corresponding list of SMILES strings and expanding R-group containing substructures to fully defined biologically relevant structures requires prior knowledge of the SPARQL query language to construct a relatively complex query as well as manual, time-consuming concatenation of results from multiple protein identifiers. P2M performs these tasks using local and optional external methods to cross-reference identifiers and identify associated chemical entities (Figure 1). To bypass the requirement of learning query languages, UniProt or EC identifiers associated with SwissProt/TrEMBL may be locally cross-referenced with the Rhea database via downloadable cross-reference mapping files. However, using downloaded data introduces the potential for the inclusion of outdated information if the most up-to-date version of a file is not used. Thus, users benefit from the ability to directly access databases rather than continually downloading fixed files.

User input to P2M can be either a list of SwissProt (a database of high quality manually annotated and reviewed protein sequences) or TrEMBL (a database of computationally annotated supplemental protein sequences) UniProt identifiers obtainable by various methods, such as alignment of protein coding sequences from an arbitrary genome to a UniProt database or download of UniProt identifiers of an organism of interest. These identifiers are used to search the Rhea database via the Rhea and UniProt SPARQL endpoints using a federated query. Additionally, the user may instead supply EC identifiers, which are cross-referenced to Rhea identifiers using the Rhea SPARQL endpoint. The result is a list of chemical structures of substrates and products associated with each Rhea identifier.

As many enzymatic reactions are defined for transforming functional groups or substructures instead of fully defined chemical structures, the resulting SMILES strings may include R-group containing compounds delineated with the "*" character. These SMILES strings are similar to the SMILES arbitrary target specification (SMARTS) language for specifying substructures, where the asterisk indicates a wildcard atom, except that the wildcards here represent any structural motif attached to the SMILES-specified structure. Thus, we term these representations as "partial structures" or "R-group representations," as the structures follow the general form *--R. Using these R-group representations can be challenging in cases where explicit compound identification is required. Thus, ChEBI identifiers of SMILES strings containing R-groups may be optionally externally queried for similar structures via Rhea/ChEBI. The similar structures option can be used to select compounds with similar chemical structures to the queried substructure and which may be fully defined in the ChEBI database. Users can enable this workflow via the “complete\_rgroups” parameter.

The output SMILES strings may be further standardized via RDKit\citep{Landrum} to strip salt ions, remove any charges, and generate canonical tautomers for each SMILES via the “clean\_smiles” parameter. Finally, lists of SMILES strings of the fully defined structures and substructures are exported to tab-delimited files (.TSV) with additional columns containing the associated ChEBI identifier and common chemical name, if available.

\section{Results and Discussion}

\begin{figure}
    \centering
    \includegraphics[width=0.9\linewidth]{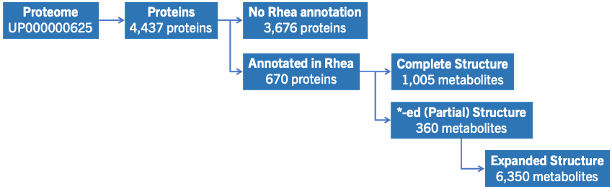}
    \caption{Summary of results for proteome-to-metabolome mapping for the annotated proteome UP000000625.}
    \label{fig:up_results}
\end{figure}

We tested our approach for identifying metabolites by applying P2M to UniProtKB identifiers extracted from the annotated proteome of E. coli strain K12 / MG1655 / ATCC 47076 (Proteome ID: UP000000625, downloaded 6/4/2021) available via the UniProt database with all P2M external querying options selected. The data consisted of 4,437 UniProtKB identifiers, from which 1,005 compounds and 360 partial structures are obtained from relevant Rhea reactions. These compounds are exported to the "smiles\_cleaned.tsv" and "smiles\_partial.tsv" output files, respectively. The partial substructures, which consist of SMILES strings containing the R-group-indicative “*” character, were expanded into 6,350 additional complete chemical structures, which are exported alongside the 1,005 other complete structures into "smiles\_expanded\_cleaned.tsv". Thus, the total metabolite set compiled by P2M consists of 7,712 unique structures as identified by ChEBI ID (Figure~\ref{fig:up_results}). Of the SMILES structures for these compounds, 7,352 (95\%) were fully defined, while 360 (5\%) depicted the partial (*-containing) substructures.

To demonstrate how P2M extracts information for specific proteins, we will examine an example protein from the E. coli strain K12 proteome, thioredoxin 2 (THIO2\_ECOLI), whose UniProt identifier is P0AGG4. The Rhea database contains information for 2 catalytic reactions involving thioredoxin 2. The output generated by P2M for P0AGG4 is shown in the tables below, omitting the "uniprot" and "smiles" columns for the complete SMILES and omitting the "uniprot" column for the partial SMILES results. The example input command, command line output, and complete results for the proteome can be found in the Supplementary Information.

\begin{table}
    \caption{Example of P2M's output for complete metabolite structures obtained for thioredoxin 2 (P0AGG4). By default, P2M will preserve information about the associated protein, annotated Rhea reaction, and compound identifiers (ChEBI ID and chemical name) for each metabolite.}
    \small
    \begin{tabular}{c c c c p{0.4\linewidth}}
        \toprule
        identifier	& chebiId	& name	   & rheaIds	& cleanedSmiles \\
        \midrule
        P0AGG4	& CHEBI:57783	& NADPH	   & 18753	        & \texttt{\tiny NC(=O)C1=CN([C@@H]2O[C@H](COP(=O)(O)OP(=O)(O)OC[C@H]3O[C@@H]( n4cnc5c(N)ncnc54)[C@H](OP(=O)(O)O)[C@@H]3O)[C@@H](O)[C@H]2O)C =CC1} \\
        P0AGG4	& CHEBI:57945	& NADH	   & 18749	        & \texttt{\tiny NC(=O)C1=CN([C@@H]2O[C@H](COP(=O)(O)OP(=O)(O)OC[C@H]3O[C@@H]( n4cnc5c(N)ncnc54)[C@H](O)[C@@H]3O)[C@@H](O)[C@H]2O)C =CC1} \\
        P0AGG4	& CHEBI:57540	& NAD(+)   & 18749	        & \texttt{\tiny NC(=O)c1ccc[n+]([C@@H]2O[C@H](COP(=O)([O-])OP(=O)(O)OC[C@H]3O [C@@H](n4cnc5c(N)ncnc54)[C@H](O)[C@@H]3O)[C@@H](O)[C@H]2O)c1} \\
        P0AGG4	& CHEBI:58349	& NADP(+)  & 18753	        & \texttt{\tiny NC(=O)c1ccc[n+]([C@@H]2O[C@H](COP(=O)(O)OP(=O)([O-])OC[C@H]3O [C@@H](n4cnc5c(N)ncnc54)[C@H](OP(=O)(O)O)[C@@H]3O)[C@@H](O)[C @H]2O)c1} \\
        P0AGG4	& CHEBI:15378	& H(+)     & 18749,18753	& \texttt{\tiny[H+]} \\
        \bottomrule
    \end{tabular}
    \label{tab:P0AGG4_nostars}
\end{table}

\begin{table}
    \caption{Example of P2M's output for partial, or *-containing, metabolite structures obtained for thioredoxin 2 (P0AGG4). The asterisks in the SMILES strings indicate wildcard structural motifs that can be expanded into structures as desired. Note that unlike the complete metabolite structure list, no "cleaned\_smiles" column exists here due to the flexible nature of the SMILES strings.}
    \small
    \begin{tabular}{c c c c c}
        \toprule
        identifier	& chebiId	& name	               & smiles	                                         & rheaIds \\
        \midrule
        P0AGG4	& CHEBI:29950	& [protein]-dithiol	    & \texttt{\tiny C(=O)(*)[C@@H](N*)CS}                   & 18749,18753 \\
        P0AGG4	& CHEBI:50058	& [protein]-disulfide	& \texttt{\tiny C([C@@H](N*)CSSC[C@@H](C(=O)*)N*)(=O)*} & 18749,18753 \\
        \bottomrule
    \end{tabular}
    \label{tab:P0AGG4_stars}
\end{table}

In the default output of P2M, all intermediate information is preserved, including the ChEBI and Rhea identifiers for metabolites associated with the given protein (Table~\ref{tab:P0AGG4_nostars}). The raw SMILES obtained via the original query is also provided in the "smiles" column (not shown), and the SMILES string following the described desalting and cleaning steps is provided in the "cleanedSmiles" columns. The user can choose to ignore the included metadata if the only data of interest are the SMILES result in the "cleanedSmiles" column.

For proteins whose reactions involve compounds with the partial SMILES structures, which represent R-group containing substructures, as components of a particular metabolic reaction, data is captured in a separate TSV file for differentiation from the complete SMILES (Table~\ref{tab:P0AGG4_stars}). The R-group expansion option in P2M, which will query ChEBI for related compounds, will expand some of these structures for inclusion into the list of expanded SMILES. However, if the user wishes to fully expand all substructures into all potential structural matches across ChEBI, P2M includes a utility for substructure expansion (\texttt{substruct2struct}) that can be used to separately expand the partial SMILES. For instance, expansion of the first substructure represented in Table~\ref{tab:P0AGG4_stars} yields 2,609 additional structures, of which a subset are shown below (Table~\ref{tab:CHEBI_29950}). We have chosen not to include full expansion via ChEBI querying due to potential strain on Rhea servers as well as limitations in utility for the end user; for instance, fully expanded substructures may not necessarily relate to the proteome in question and further downselection is thus required on the user's end. However, a tutorial on using \texttt{substruct2struct} is included in the GitHub repository for interested users.

\begin{table}
    \caption{Output showing the first 10 compounds (out of 2,609) obtained by applying substructure expansion of [protein]-dithiol (CHEBI:29950) from the partial structure associated with thioredoxin 2 (P0AGG4) in Table~\ref{tab:P0AGG4_stars}. Structures were obtained via the \texttt{p2m.query.substruct2struct} utility. The full list of structures is included in the Supplementary Information}
    \small
    \begin{tabular}{c c p{0.5\linewidth}}
        \toprule
        chebiId	     & name	               & rheaIds \\
        \midrule
        CHEBI:159602 & Asn-Asp-Cys	       & \texttt{\tiny SC[C@H](NC(=O)[C@@H](NC(=O)[C@@H](N)CC(=O)N)CC(O)=O)C(O)=O} \\
        CHEBI:39390	 & felinine            & \texttt{\tiny CC(C)(CCO)SC[C@H](N)C(O)=O} \\
        CHEBI:53600	 & benzylpenicillanyl group	& \texttt{\tiny [H][C@]12SC(C)(C)[C@@H](N1C(=O)[C@H]2NC(=O)CC3=CC=CC=C3)C(*)=O} \\
        CHEBI:135919 & nesiritide	       & \texttt{\tiny [H]N[C@@H](CO)C(N1CCC[C@H]1C(N[C@@H](CCCCN)C(N[C@@H](CCSC)C(N[C@@H](C(C)C)C (N[C@@H](CCC(N)=O)C(NCC(N[C@@H](CO)C(NCC(N[C@H]2CSSC[C@H](NC(CNC([C@H](CC(C) C)NC(CNC([C@@H](NC([C@@H](NC([C@@H](NC([C@@H](NC([C@H]([C@H](CC)C)NC([C@@H]( NC([C@@H](NC([C@H](CCSC)NC([C@H](CCCCN)NC([C@@H](NC(CNC([C@@H](NC2=O)CC3=CC= CC=C3)=O)=O)CCCNC(N)=N)=O)=O)=O)CC(O)=O)=O)CCCNC(N)=N)=O)=O)CO)=O)CO)=O)CO)= O)CO)=O)=O)=O)=O)C(N[C@@H](CCCCN)C(N[C@@H](C(C)C)C(N[C@@H](CC(C)C)C(N[C@@H]( CCCNC(N)=N)C(N[C@@H](CCCNC(N)=N)C(N[C@@H](CC4=CNC=N4)C(O)=O)=O)=O)=O)=O)=O)= O)=O)=O)=O)=O)=O)=O)=O)=O)=O} \\
        CHEBI:159670 & Asn-Cys-Tyr	       & \texttt{\tiny SC[C@H](NC(=O)[C@@H](N)CC(=O)N)C(=O)N[C@@H](CC1=CC=C(O)C=C1)C(O)=O} \\
        CHEBI:163589 & Gly-Cys-Phe	       & \texttt{\tiny SC[C@H](NC(=O)CN)C(=O)N[C@@H](CC1=CC=CC=C1)C(O)=O} \\
        CHEBI:160245 & Asn-Val-Cys	       & \texttt{\tiny SC[C@H](NC(=O)[C@@H](NC(=O)[C@@H](N)CC(=O)N)C(C)C)C(O)=O} \\
        CHEBI:139364 & piperacilloyl-L-lysine   & \texttt{\tiny N1[C@H](C(S[C@]1([H])[C@H](NC([C@@H](C2=CC=CC=C2)NC(N3CC(NC(C3=O)=O)CC)=O)= O)C(=O)NCCCC[C@@H](C(=O)O)N)(C)C)C(O)=O} \\
        CHEBI:161744 & Cys-Trp-Asn	       & \texttt{\tiny SC[C@H](N)C(=O)N[C@@H](CC=1C=2C(NC1)=CC=CC2)C(=O)N[C@@H](CC(=O)N)C(O)=O} \\
        CHEBI:157796 & Asn-Cys	           & \texttt{\tiny SC[C@H](NC(=O)[C@@H](N)CC(=O)N)C(O)=O} \\
        ... & ... & ... \\
        \bottomrule
    \end{tabular}
    \label{tab:CHEBI_29950}
\end{table}

\section{Conclusion}
P2M is a software tool for identifying SMILES strings of enzymatic reactants and substrates given UniProt or EC protein identifiers via a user-friendly command line interface. We anticipate its broader use toward building identification libraries for metabolites likely to be observed in a biological system. In the context of complex mass spectrometry sample analysis, P2M can be used to limit the initial chemical search space, thus reducing misidentifications and computation time. P2M is open source and available to the public free of charge. For installation instructions and source code, see https://github.com/pnnl/p2m.

\section{Data and Software Availability}

P2M is written in Python and is available as an open sourcee, publicly available software package via GitHub at https://github.com/pnnl/p2m. The UniProt input data and generated output files are available to download as part of the Supplementary Files.

\begin{acknowledgement}

    This work was supported by the PNNL Laboratory Directed Research and Development program and is a contribution of the \textit{m/q} Initiative. PNNL is operated by Battelle for the DOE under contract DE-AC05-76RL01830.
    
\end{acknowledgement}

\begin{suppinfo}

A listing of the contents of each file supplied as Supporting Information
should be included. For instructions on what should be included in the
Supporting Information as well as how to prepare this material for
publications, refer to the journal's Instructions for Authors.

The following files are available free of charge.
\begin{itemize}
  \item SMILES.csv: Table containing information on the molecules listed in the example in the manuscript for thioredoxin 2 (P0AGG4), including SMILES and associated identifiers.
  \item example\_run.zip: Zipped folder containing an example run of P2M. Files include an example input file (ecoli\_k12\_mg1655\_uniprotkb.txt), output files, and command line log (output.log).
  \item CHEBI\_29950.csv: Results from the \texttt{substruct2struct} expansion for CHEBI:29950.
\end{itemize}

\end{suppinfo}

\section{Author Contributions}

Author contributions are listed using  CRediT Contributor Roles Taxonomy (CRediT) descriptors (https://credit.niso.org/). C.H.C contributed via software, project administration, and writing - review \& editing tasks. B.J.K contributed via conceptualization, data curation, methodology, software, and writing - original draft tasks. R.S.R contributed via conceptualization, project administration, and writing - review \& editing tasks. S.M.C contributed via supervision and writing - review \& editing tasks.

\bibliography{main}

\end{document}